\documentstyle[12pt]{article}
\begin{document}

\title{ About the modern "experimental value" of $W$ boson width.}

\author{I.F.Ginzburg\\Laboratoire de Physique Th\'eorique
EN{\large S}{\Large L}{\large A}PP, France\thanks{ B.P.110, 74941
Annecy-Le-Vieux Cedex, France},\\ Institute of Mathematics,
Novosibirsk, Russia\thanks{Permanent address: Institute of
Mathematics, 630090, Novosibirsk, Russia; e-mail:
ginzburg@math.nsk.su}\\ and\\K.Melnikov\\Institute fur
Physik,Universit\"{a}t Mainz\thanks{ D 55099 Germany, Mainz, Johannes
Guttenberg Universit\"{a}t, Institute f\"{u}r Physik, THEP, Staudinger
weg 7.}\\ \\MZ-TH/95-03\\ENSLAPP-A-505/95}

\maketitle

\begin{abstract}
It is shown that the methods which have been used up to now to
determine the $W$ width from the $p\bar p$ data confirm the  SM
predictions for some
combinations of various
phenomenological parameters, however, they do not give an independent value
for the $W$ width.  Moreover,
 the accuracy that could be achieved in future
experimental checks of SM predictions for such quantities
is limited by effects which require detailed theoretical study.
\end{abstract}

{\large\bf Introduction}\\[0.1cm]

Recent results from LEP and SLC
have given us a value for  the mass and the
width of Z boson with a spectacular precision. The same problem
for W boson is studied at the Fermilab $p\bar p$ collider.

In this note, we discuss experimental results related to the W--boson decay
width $\Gamma_W$. These have been obtained by CDF and D0
collaborations by two methods -- the "indirect" (see \cite{appr}) and
the "direct" one \cite{Abe},\cite{Rosner}.
These results are in agreement with each other. They confirm
the Standard Model (SM)
with 3 families including a top quark much heavier than the $W$ boson. The
value of $\Gamma_W$ obtained in these measurements is quoted
now in Particle Data Review \cite{PDG}.

Even if we assume that our
knowledge of the quark distribution functions in the proton is
precise enough, the results of the
 experiments just mentioned  are described by some complex relations,
containing both $\Gamma_W$ and other quantities, which can be
determined only by the use of the SM, either explicitly or implicitly.
Hence, these experiments check SM predictions in this
%complex
form only.

{\em Consequently, the results of these experiments can not be treated as
the independent experimental value of the W boson width.}

Due to the fact that {\em (i)} the results of these experiments are
expressed through a number of phenomenological parameters of the SM and
{\em (ii)} New Physic can manifest itself in various ways, i.e. influence all
the
SM parameters used for the width extraction from the data, the analyses
of possible New Physic manifestation in these experiments by looking for the
deviations from the SM  in $\Gamma _W$ {\em only} seem
to be meaningless.
 Besides, possible
improvement of the accuracy of such SM confirmation is
limited by our insufficient knowledge of
the parton distributions in the proton and by some nontrivial
radiative corrections.

To explain these statements, we {\em (i)} briefly reproduce ideas
of the methods \cite{appr,Abe} without making any criticism ; {\em
(ii)} consider the  real relation between the data obtained
and  the SM; {\em (iii)}
briefly
discuss difficulties with the possible improvement of the confirmation
of the SM within these methods. \\[0.1cm]

{\large\bf Main points of the experimental methods.}\\[0.1cm]

In both methods, $W$ production is recorded as an event with
production of a  lepton (for example, electron) having high
transverse momentum. A large transverse energy imbalance is
required to signal the presence of a neutrino.\\[0.1cm]

{\large\em ``Indirect" method.}

This is the ``eldest" method which has been used to obtain
$\Gamma_W$ from the data.  The value of $\Gamma _W$ obtained by
 this method is quoted now in \cite{PDG}.  One can find
a detailed description in ref. \cite{appr}.

In this method, the  experiment gives the total number of
``real" $W$'s produced which have then decayed
into  e.g. $e\nu $ (with necessary cuts). The
number of events is written as the product of the
$W$ boson production cross section $\sigma (W)$, the corresponding
branching ratio and total luminosity $L$:
\begin{equation}
N_{W/e}=L\cdot\sigma (W)\cdot Br(W\to e\bar\nu) \label{event}
\end{equation}

The production cross section $\sigma (W)$ is calculated using
known structure functions in the standard way, assuming that the
coupling of the quarks to the $W$ boson is given by the SM.

The production cross section $\sigma(W)$ and the luminosity $L$
are known with bad accuracy.
%To avoid this inaccuracy, the $e^-e^+ \;(Z)$ production with high enough
%%transverse momentum of
%leptons is considered additionally
To circumvent this problem, the production of
$e^-e^+$ pairs  with high transverse momentum is also considered,
these events occur through the production and subsequent decay of Z-boson.
 The number of events of this type is written similarly:
$$
N_{Z/e}=L\cdot\sigma (Z)\cdot Br(Z\to e^-e^+).
$$

The observed ratio of $N_{W/e}$ and $N_{Z/e}$
\begin{equation}
\frac{N_{W/e}}{N_{Z/e}}=\frac{\sigma (W)}{\sigma (Z)}\cdot
\frac{Br(W\to e\bar\nu)}{Br(Z\to e^-e^+)}\label{ratio}
\end{equation}
is free from a number of inaccuracies which are inherent to both
quantities $N_{W/e}$ and $N_{Z/e}$ if they are treated
separately.
%Indeed, badly defined factor $L$ is absent here.
Indeed, the poorly known luminosity factor, $L$, has dropped out.
Second, the ratio of production cross sections is calculated with
high accuracy because these cross
sections are defined by the same structure function. The effect of
radiative corrections to these
cross sections is beyond the accuracy of this calculation.

Since  $Br(Z\to e^-e^+)$ is known precisely from the LEP data,
 the above ratio gives $Br(W\to e\bar\nu)$. However,  in
order to extract the value of the $W$ width from the data one
needs extra input. The assumption that the partial width
$\Gamma(W\to e \bar\nu)$ is just given by the SM is used for this
aim. Finally,
\begin{equation}
\Gamma_W =\Phi  \frac{N_{Z/e}}{N_{W/e}}; \quad \Phi=
\frac{\Gamma(W\to e\bar\nu)}{Br(Z\to e^+e^-)}\Sigma(W/Z);\quad
\Sigma(W/Z)=\frac{\sigma (W)}{\sigma (Z)}.\label{Br}
\end{equation}

%Let us write in more details the factor $\Sigma(W/Z)$:

In more details, the factor $\Sigma(W/Z)$ is written:
\begin{equation}
\Sigma(W/Z) \propto \frac{\nu(u\bar d/u)\Gamma(W\to u\bar d)+\nu(c\bar
s/u)\Gamma(W\to c\bar s)} {\sum\limits_{q=u,d,s,c,b}
\nu( q/u)\Gamma(Z\to q\bar q)}.\label{Phi}
\end{equation}
Here quantities $\nu$ are expressed through the quarks and
antiquarks distribution functions in the proton, for instance,
\begin{equation}
\nu(u\bar d/u) =\frac {<n_u(x_1)n_{\bar d}(x_2)|_{x_1
x_2=M_W^2/s}>}{ <n_u(x_1)n_{\bar u}(x_2)|_{x_1
x_2=M_Z^2/s}>},...;\;(\nu(u /u)=1),\label{nu}
\end{equation}
where $<>$ means  averaging with the use of the experimental cuts.

Some additional assumptions make possible further simplifications
for the ratio of the cross
sections $\Sigma(W/Z)$. For example, one can neglect the
contribution of charmed quarks.
In this case,  the numerator of this
ratio contains  only the first term, while  the denominator has three
terms, which correspond to $u,d,s$ quarks.
The quantity $\Gamma(W\to u\bar d)$ has to be calculated within the SM
together with $\Gamma(W\to e\bar \nu)$.

Since the factor $\Phi$ is calculated with good precision,
equation (\ref{Br}) give us the "experimental" value of
$\Gamma_W$.\\[0.1cm]

{\large\em "Direct" method.}

Another approach to the $\Gamma_W$ measurement has been
proposed in ref. \cite{Rosner}. This method with little
modifications has been recently used by CDF group \cite{Abe}.

The idea is to study the production of $e\nu$ system, with an
invariant mass $Q$ which is larger than some value $Q_0\gg M_W$
and to compare it with $W (e\nu$) production, described by eq.
(\ref{event})\footnote{ In the actual experiment
\cite{Abe} the transverse mass of the $e\nu$ system is used
rather then the invariant mass. A cut in the transverse mass
$M_{\bot} > 110$ GeV is imposed.}. It is assumed that all these
events are generated
via the production of highly virtual $W$ bosons. The number
of events is given by the approximate equation (similar to the
equation, proposed in ref.
\cite{berends} for the {\em narrow} region near Z pole):
\begin{equation}
N(Q)\propto L\cdot\int\limits_{Q^2>Q^2_0} dQ^2
\sigma(W,Q)\frac{ Q^2\Gamma(W\to e\nu)}
{(Q^2-M_W^2)^2+\Gamma_W^2 \cdot (Q^2)^2/M_W^2},\label{eventQ}
\end{equation}
where integration over other parameters with suitable kinematical cuts
is assumed.

In this equation $\sigma(W,Q)$ stands for the production cross
section of the off-shell
$W$ and a specific form of $Q^2$ dependence for  the W
width (both total and partial) is assumed. To calculate
$\sigma(W,Q)$ the same approximation
for partial decay widths of the $W$ to quarks is used and the
convolution of the distribution
functions in the new point $Q^2$ is evaluated.

Then, similarly to the "indirect" method, one considers the ratio of
the quantities described by eq. (\ref{eventQ}) and eq.
(\ref{event}). Since $\Gamma(W\to e\bar \nu)/Br(W\to e\bar \nu) =
\Gamma_W$, this new ratio is written in the form
\begin{equation}
\frac{N(Q)}{N(W/e)} \propto  \int\limits_{Q^2>Q^2_0} dQ^2
\frac{Q^2\Gamma_W\Sigma(Q)}{[(Q^2-M_W^2)^2+Q^4\Gamma_W^2/ M_W^2] M_W};\;
\Sigma(Q)=\frac{\sigma(W,Q)}{\sigma (W)}.\label{ratio2}
\end{equation}
Here, the factor $\Sigma(Q)$ is calculated with the same (or better)
accuracy as in ratio (\ref{ratio}). Indeed, just as for the
``indirect'' case, we have
\begin{equation}
\Sigma(Q)=\frac{\nu(u\bar d/u;Q)\Gamma(W\to u\bar d;Q)+\nu(c\bar
s/u;Q)\Gamma(W\to c\bar s;Q)}{\nu(u\bar d/u)\Gamma(W\to u\bar d)+\nu(c\bar
s/u)\Gamma(W\to c\bar s)} .\label{ratio3}
\end{equation}
The new notations are evident from a comparison with eq. (\ref{nu}).

Using the extrapolation for partial widths in the spirit of
ref. \cite{berends} and neglecting $c$ quark content in the
proton, this quantity transforms into the ratio of quark numbers at
different $x$, and the final equation (\ref{ratio2})
% has no terms, calculated in the SM.
does not contain any term  calculated in the SM.

Hence, the ratio of events (\ref{ratio2}) depends only on  one unknown
quantity: the total width $\Gamma_W$. Therefore, the value of
$\Gamma _W$ is obtained by fitting
the data with this equation.\\[0.1cm]

{\large\bf Relation to the Standard Model and the effects of
New Physics in  experiments.}\\[0.1cm]

{\em Indirect method.} The analysis after eq. (\ref{Br}) shows that
the SM
has been used repeatedly for the calculation of the quantity $\Phi$ in
the right hand side
of this equation. It is necessary to calculate
both  $\Gamma(W\to u\bar d)$ and
 $\Gamma(W\to e\bar \nu)$, however
 these calculations have
 the same status as the calculation of $\Gamma_W$.
They rely on an assumption
 about the existence of three families
with a very heavy t--quark
\footnote{ Certainly, at the modern level of the SM
verification the leptonic widths have been calculated in tree
approximation of SM, while for the quark widths one--loop gluon
corrections \cite{RC} have been taken into account.}.
 Moreover,
the partial widths of $Z$ decay into various light quark systems have
not
been measured separately. Hence, they are calculated within
the SM only. Therefore, the
indirect method gives some combination of various
phenomenological parameters, but not $\Gamma_W$ separately.

{\em Direct method}.
At first glance, we deal here with a  much better situation.
Indeed, SM calculations in this case
have  dropped out from the ratio of the cross sections $\Sigma(Q)$.
Unfortunately, this conclusion is inexact. Indeed, the crucial
point of this method is the use of the W propagator in the specific form
(\ref{eventQ}) and the corresponding extrapolation for W partial
widths. The  equation used  for the propagator
has been proposed in ref. \cite{berends} as an
approximation which is valid near the W pole only. It was not proven for
the case $Q^2\gg M_W^2$ discussed here. Hence, the basic equation
(\ref{eventQ}) above is unfounded.
To obtain the correct form of the corresponding cross section,
the radiative corrections should be taken into account both to the W
propagator itself (real part of its polarization operator) and to the
partial widths of W decay to leptons or quarks, which are latent in
the final result. In particular, new channels
(like $W^*\to t\bar b,\; W^*\to W\gamma$)
contribute more and more
strongly to the effective total W width $\Gamma_W(Q)$
with the growth of $Q$.

 Even if one would take these points into account, the basic problem
will still be there: just as in the indirect method we use SM predictions
for some basic quantities in the equations. Therefore, the  "direct" method
gives us in fact some complex object (which {\em as a whole} can be
predicted by the SM) but not the value of $\Gamma_W$ separately.

{\em Relation to the New Physics Effects.}
The main goal of similar work is to look for possible
deviations from the SM -- effects of the New Physics. These effects
can show up in various ways, i.e. they can change  all quantities used for the
description of the  experiments discussed (the modifications in
$\Gamma _W$ is only one possibility with a lot of the others
being neglected in the basic equations due to the use of the SM).

To make this point more clear, we present some partial list of
opportunities. Perhaps, some of them are excluded by other data,
but in each case special study is needed in order to ignore
a particular model  in the analysis of the  experiments discussed.

For example, one can
imagine that there is some small additional fraction of observed
high $p_{\bot}$ leptons due to their production in the decay of
selectron or smuon or excited electron or muon or any other particle with
high enough mass, which can be produced either directly (via
photon or Z) or through W decay. Besides, some new thresholds
could be opened with the increase in the effective mass $Q$ of the
produced $W\to e\nu$ system (both standard ($t\bar b$) and
"non--standard" channels). They can increase total W width and
decrease visible leptonic Branching Ratio. This effect is particularly
dangerous in the "direct" method. Similar effects can be
connected with the admixture of additional heavy $W$ bosons (from some
extension of SM).\\[0.1cm]

{\large\bf The accuracy of possible forthcoming confirmations of
the SM in these methods.}\\[0.1cm]

It seems that the experiments which have been
discussed so far provide a good place for the test
of the SM predictions for the ratios of the  number of events. For example,
CDF group believes that in the framework of the ``direct'' method
``... with future runs of the Fermilab collider, a 30 MeV
measurement (of $W$ width) is possible which approaches the level
of the radiative corrections to the width.''\cite{Abe}.

The statement about the  measurement of $\Gamma_W$ has already  been discussed
above, but the aim to achieve an accuracy of $\sim 1\%$ in these experiments
(which is indeed the level of the SM electroweak
corrections) introduces further problems.
%but the authors believe, in fact,  that the accuracy $\sim 1 \%$
%in these
%experiments can be achieved (this is indeed the level of the SM electroweak
%corrections).
Unfortunately, there is no theory that can describe
the data with such accuracy even within the SM. Let us discuss briefly
the difficulties associated with the proposed
level of accuracy.

First of all, the W transverse momentum distribution enters
the actual data analyses \cite{Abe}. The contribution
of the region of small $p_{\bot}$ in  this distribution is very important,
but now it can only be obtained with poor accuracy, especially for
$p_{\bot}\le 10 $ GeV. This leads to an uncertainty in the final
result which has been estimated as  $\sim 2\%$
 for the  ``indirect'' method \cite{ratio}.
 For the ``direct" method this uncertainty has not
been discussed yet. Besides,  gluon radiation
and processes like $s+g\to W+c$ should vary the
W distribution over $p_{\bot}$ with an increase in  $Q$.
These effects should be taken into account in the precise analysis
of the results obtained by  the
direct method. Finally, inaccuracy due to the ignorance
of the c quark contribution should be estimated too.

Let us assume however that this difficulty can be overcome. If so,
 more delicate questions, connected with the calculation of
radiative corrections to the basic process $p\bar p \to W +...\to
e\nu+...$, become important and require additional theoretical
work. Let us mention only two of them.

{\em (i)}. With higher accuracy, simple Breit--Wigner description
of the unstable gauge boson propagators becomes inadequate for
different reasons. For example, the difficulties connected with
the gauge invariance have been recently pointed out for $e^+e^-$
and $\gamma\gamma$ collisions \cite{ACO}. Hence, methods like
those developed in refs. \cite{Stuart} have to be used at least.

{\em (ii)}. QCD radiative corrections to the $W$ production
process give here the $Q^2$ dependent $K$ -- factors (similar to
the standard Drell--Yan process description). The new point here is
the fact that the electromagnetic corrections should be taken
into account in these $K$ -- factors since W boson is the charged
particle in contrast to the photon. \\[0.1cm]

Therefore, {\em the experimental value for the width of the W
boson $\Gamma_W$ is absent now}. We don't see any method for the
determination of this width before LEP2 operations.

Nevertheless, the expected precision
in the  confirmation of the SM in the  experiments discussed
is remarkable. Perhaps, it will be useful to consider them
as new experiments for testing  QCD and proton structure. \\[0.1cm]

We are grateful to G.B\'elanger, F.Berends, E.Boos, F.Boudjema,
F.Cuypers, S.Eidelman, V.Ilyin, K.Kato, J.Kurihara, G.Oldenborgh,
V.Serbo, D.Schildknecht for discussions.

The work of I.F.G. is supported by grants ISF RPL000 and INTAS -- 93 --
1180. K.M. is grateful to the Graduiertenkolleg Teilchenphysik,
Universit\"{a}t Mainz for support.

\end{document}